\documentclass[conference]{IEEEtran}
\usepackage{graphicx} 
\usepackage{cite} 
\usepackage{amsmath,amssymb} 
\usepackage{longtable}
\usepackage{tabularx}
\usepackage{booktabs}
\usepackage[utf8]{inputenc}
\usepackage{fancyhdr}
\usepackage{lipsum} 

\title{Virtual Reality and Augmented Reality Security: A Reconnaissance and Vulnerability Assessment Approach}
\author{
    \IEEEauthorblockN{Sarina Dastgerdy}
    \IEEEauthorblockA{
        \textit{Department of Computer Science}\\
        \textit{University of Guelph}\\
        Guelph, Canada \\
        Email: sdastger@uoguelph.ca
    }
}
\date{May 2024}

\begin{document}

\maketitle

\begin{abstract}
Various industries have widely adopted Virtual Reality (VR) and Augmented Reality (AR) technologies to enhance productivity and user experiences. However, their integration introduces significant security challenges. This systematic literature review focuses on identifying devices used in AR and VR technologies and specifies the associated vulnerabilities, particularly during the reconnaissance phase and vulnerability assessment, which are critical steps in penetration testing.
Following Kitchenham and Charters' guidelines \cite{Guidelines}, we systematically selected and analyzed primary studies. The reconnaissance phase involves gathering detailed information about AR and VR systems to identify potential attack vectors. In the vulnerability assessment phase, these vectors are analyzed to pinpoint weaknesses that malicious actors could exploit.
Our findings reveal that AR and VR devices, such as headsets (e.g., HTC Vive, Oculus Quest), development platforms (e.g., Unity Framework, Google Cardboard SDK), and applications (e.g., Bigscreen VR, VRChat), are susceptible to various attacks, including remote code execution, cross-site scripting (XSS), eavesdropping, and man-in-the-room attacks. Specifically, the Bigscreen VR application exhibited severe vulnerabilities like remote code execution (RCE) via the 'Application.OpenURL' API, XSS in user inputs, and botnet propagation \cite{vondravcek2023rise}. Similarly, the Oculus Quest demonstrated susceptibility to side-channel attacks and ransomware \cite{silva2023survey}.
This paper provides a detailed overview of specific device vulnerabilities and emphasizes the importance of the initial steps in penetration testing to identify security weaknesses in AR and VR systems. By highlighting these vulnerabilities, we aim to assist researchers in exploring and mitigating these security challenges, ensuring the safe deployment and use of AR and VR technologies across various sectors.

\end{abstract}

\section{Introduction}
Virtual Reality (VR) involves creating a realistic environment using computer graphics. Unlike static scenes, this artificial world dynamically responds to user interactions, emphasizing the essential aspect of real-time interactivity. In this setting, real-time indicates that sensors detect user inputs, allowing the computer to instantly alter the virtual environment \cite{burdea2003virtual, 1}. Augmented Reality (AR), on the other hand, overlays virtual computer-generated information onto a real-time view of the physical world, whether viewed directly or indirectly, enhancing the user's perception of their environment \cite{furht2011handbook,2}.\\
The development of VR and AR technologies has a rich history, beginning with early experimental systems in the 1960s and 1970s. Humans have long sought to bring imagined worlds into reality, from art and literature to modern digital media. In the 1980s, pioneers like Jaron Lanier popularized "virtual reality," leading to the first VR systems such as NASA Ames' VIEW lab, which featured advanced displays and gesture tracking \cite{historybryson2013virtual}. Milestones like the Oculus Rift in 2012 and Microsoft HoloLens in 2015 brought VR and AR into mainstream use, revolutionizing gaming and expanding AR applications. Subsequent devices like the HTC Vive and Google Cardboard made VR and AR more accessible. Companies like Oculus, Samsung, and Microsoft have been crucial in advancing these technologies, enhancing interactivity and presence. This evolution has moved us closer to the Ultimate Display—a device that mimics reality perfectly \cite{historybown2017looking,3}. The historical progression of VR and AR technologies highlights their transformative potential and growing impact on various aspects of modern life.\\
Recent technological breakthroughs have significantly increased the application of AR and VR in numerous industries. These technologies improve maintenance, repair, and assembly activities in sectors such as aerospace, military, automotive, construction, and healthcare, boosting productivity and lowering risk. In education and healthcare, they produce realistic training spaces that save costs and hazards. AR and VR also improve client interactions in industries like automotive, banking, consumer products, media, entertainment, and tourism, resulting in increased customer engagement and revenue. Furthermore, they aid in design and data analysis, which benefits industries such as aerospace, automotive, and construction by enhancing efficiency and detecting design flaws early \cite{ivanova2018vr,4}. Given their widespread use and critical role across various sectors, protecting AR and VR systems is paramount. However, the integration of these technologies introduces significant security challenges that must be addressed to safeguard end users. A thorough risk analysis reveals that the impact of a security breach can vary widely depending on the application domain. Some of the most critical attacks include overlay, human joystick, and tampering, with the gaming and educational sectors being particularly vulnerable. In the most severe cases, these attacks can lead to extreme consequences such as the death of individuals or the bankruptcy of organizations. Given these risks, it is imperative to identify and mitigate security vulnerabilities in VR and AR systems through rigorous penetration testing \cite{silva2023survey,1h,5}.\\
To address security issues and adhere to mandated regulations, security experts have developed several assurance methods, including vulnerability detection, proof of correctness, layered design, and penetration testing. Penetration testing assesses the difficulty an attacker would have in breaching an organization’s security controls. This process involves simulating unauthorized attacks using automated tools, manual techniques, or both, to identify vulnerabilities in a controlled environment. The primary goal of penetration testing is to uncover security weaknesses so they can be addressed before malicious actors exploit them \cite{shah2015overview}. Penetration testing consists of three main phases: preparation, implementation, and analysis. This paper focuses on the implementation phase, specifically on information gathering and vulnerability analysis. In the information gathering step, testers scan and identify all logical and physical areas to collect necessary data. Based on this information, they analyze and assess existing vulnerabilities \cite{al2018study}. 
\subsection{Prior Research}
Regarding penetration testing approaches for AR and VR systems, there is a noticeable scarcity of Systematic Literature Reviews (SLRs). A notable recent survey, titled "A Systematic Threat Analysis and Defense Strategies for the Metaverse and Extended Reality Systems," provides a comprehensive overview of AR and VR applications. This paper identifies the vulnerabilities and potential cyber threats associated with Extended Reality (XR) and metaverse technologies, including risks to human health, psychology, and monetary fraud. The detailed taxonomies of XR cyber threats and corresponding defensive strategies offer valuable insights for researchers, developers, and policymakers aiming to mitigate these risks. However, this survey does not address vulnerabilities related to specific devices or the reconnaissance phase of penetration testing \cite{qamar2023systematic}.\\
Another recent study, titled "A Survey and Risk Assessment on Virtual and Augmented Reality Cyberattacks," focuses on risk assessment. This research identifies 15 types of attacks and provides a risk analysis based on the sector affected, the likelihood of each attack, and its potential impact. Unlike the previous paper, this study assesses risk across various devices. However, it does not delve into identifying specific vulnerabilities \cite{silva2023survey,6}.\\
\subsection{Research Goals, Contributions, and Layout}
The aim of this paper is to provide a comprehensive overview of the research on various devices used in AR and VR technologies, along with the specific vulnerabilities identified for each device. This comprehensive data is intended to assist other researchers in further expanding existing research or exploring other aspects of penetration testing. The primary research questions addressed are:
\begin{itemize}
\item \textit{\textbf{RQ1:} What devices are used to implement AR and VR technology?}
\item \textit{\textbf{RQ2:} What vulnerabilities are associated with these specific devices?}
\end{itemize}
These questions guide the discussion to the reconnaissance phase and involve gathering detailed information on all known vulnerabilities. The goal is to provide researchers with a concise and updated overview of these critical steps, enabling them to address gaps or utilize this information for subsequent phases of penetration testing.\\
The key contributions of this paper are as follows:
\begin{itemize}
\item A systematic review of AR and VR devices and their associated vulnerabilities, providing an updated and comprehensive dataset for further research.
\item Identification and categorization of specific vulnerabilities based on AR and VR technologies.
\item Emphasis on the importance of the reconnaissance phase and vulnerability assessment in penetration testing, offering a foundation for more in-depth exploration of AR and VR security challenges.
\item Provision of actionable insights and a structured framework for researchers to build upon, facilitating the development of enhanced security measures for AR and VR systems.
\end{itemize}

The structure of this paper is as follows: In Section 2, the methodology for systematically selecting the primary studies is detailed. Section 3 provides an overview of the results obtained from these selected studies. Section 4 delves into an analysis of the findings with respect to the research questions introduced earlier. Finally, Section 5 summarizes the research, offering conclusions and proposing directions for future studies.

\section{Research Methodology}
We carried out the Systematic Literature Review (SLR), following the guidelines provided by Kitchenham and Charters \cite{Guidelines} to accomplish the goal of responding to the research questions. Three main steps are defined in SLR: the planning, conducting, and reporting stages of the review in loops.

\subsection{Selection of Primary Studies}
To select primary studies, keywords with boolean operators were applied in search engines. The search strings used to find relatable literature are: 
("virtual reality" OR "VR" OR "augmented reality" OR "AR") AND ("security" OR "cyber-security" OR "penetration testing" OR "security testing" OR "vulnerability assessment" OR "threat assessment")
Searches conducted from 27th May until 12th June, among well-known journal databases mentioned below:
- IEEE Xplore\\
- ScienceDirect\\
All papers from 2020 to 2024 from these specified sources were reviewed and considered for inclusion based on their relevance to the research questions.

\subsection{Inclusion and Exclusion Criteria}
To perform the best SLR research and answer the questions, papers included should pass a filtering process. To do so, inclusion and exclusion criteria are specified in Table \ref{tab:criteria}.
\begin{table}[ht]
    \centering
    \caption{Inclusion and Exclusion Criteria}
    \begin{tabular}{|>{\centering\arraybackslash}m{4cm}|>{\centering\arraybackslash}m{4cm}|}
        \hline
        \textbf{Inclusion Criteria} & \textbf{Exclusion Criteria} \\
        \hline
        \begin{itemize}
            \setlength\itemindent{-1.5em} 
            \item[-] The article must be published in a peer-reviewed source
            \item[-] The article must be written in English
            \item[-] The article must be accessible
            \item[-] The article must explore at least one stage of penetration testing in VR and AR environments.
        \end{itemize} &
        \begin{itemize}
        \setlength\itemindent{-1.5em} 
            \item[-] Articles that do not discuss security concerns, vulnerabilities, threats, or risks associated with VR or AR environments.
            \item[-] Non-English articles
            \item[-] Books
            \item[-] White papers
        \end{itemize} \\
        \hline
    \end{tabular}
    \label{tab:criteria}
\end{table}

\subsection{Selection Results}
Based on the research keywords, 13 papers were found from IEEE Xplore from 2020, of which 12 were accessible. On the ScienceDirect website, papers were listed in order of relevance; however, based on the inclusion and exclusion criteria, only 10 were selected.

\subsection{Quality Assessment}
The quality assessment of the selected studies was conducted based on the following aspects:
\begin{itemize}
    \item \textbf{Relevance}: The study's relevance to the research questions.
    \item \textbf{Methodology}: The robustness and appropriateness of the research design and methods used.
    \item \textbf{Findings}: The clarity and significance of the study's findings and conclusions.
    \item \textbf{Publication Quality}: The quality of the publication venue, such as peer-reviewed journals or conferences.
\end{itemize}
Each criterion was rated on a scale of low, medium, and high to ensure a comprehensive evaluation. Studies had to meet a minimum threshold of at least one medium rating and three high ratings to be included in the final analysis. After the final assessment, only 5 out of the 22 primary papers reached the threshold for inclusion.

\subsection{Data Extraction}
Data were extracted from all papers that passed the quality assessment to ensure the accuracy and consistency of the review. The extraction focused on capturing the most pertinent information, categorized into the following sections:
\begin{itemize}
\item \textbf{AR or VR Focus}: Identifying whether the study focused on AR or VR technologies.
\item \textbf{Attack or Vulnerability Assessment}: Identifying if the paper assessed attacks or vulnerabilities.
\item \textbf{Focus Area}: Identifying if the paper assessed physical devices, applications, software, environments, or development platforms.
\end{itemize}
The extracted data were documented and stored in a spreadsheet to facilitate easy retrieval and analysis.

\subsection{Data Analysis}
To answer our research questions, data were extracted and supplemented with a meta-analysis to reach more robust conclusions.
\subsubsection{Publications Over Time}
\begin{figure}[ht]
  \centering
  \includegraphics[width=\linewidth]{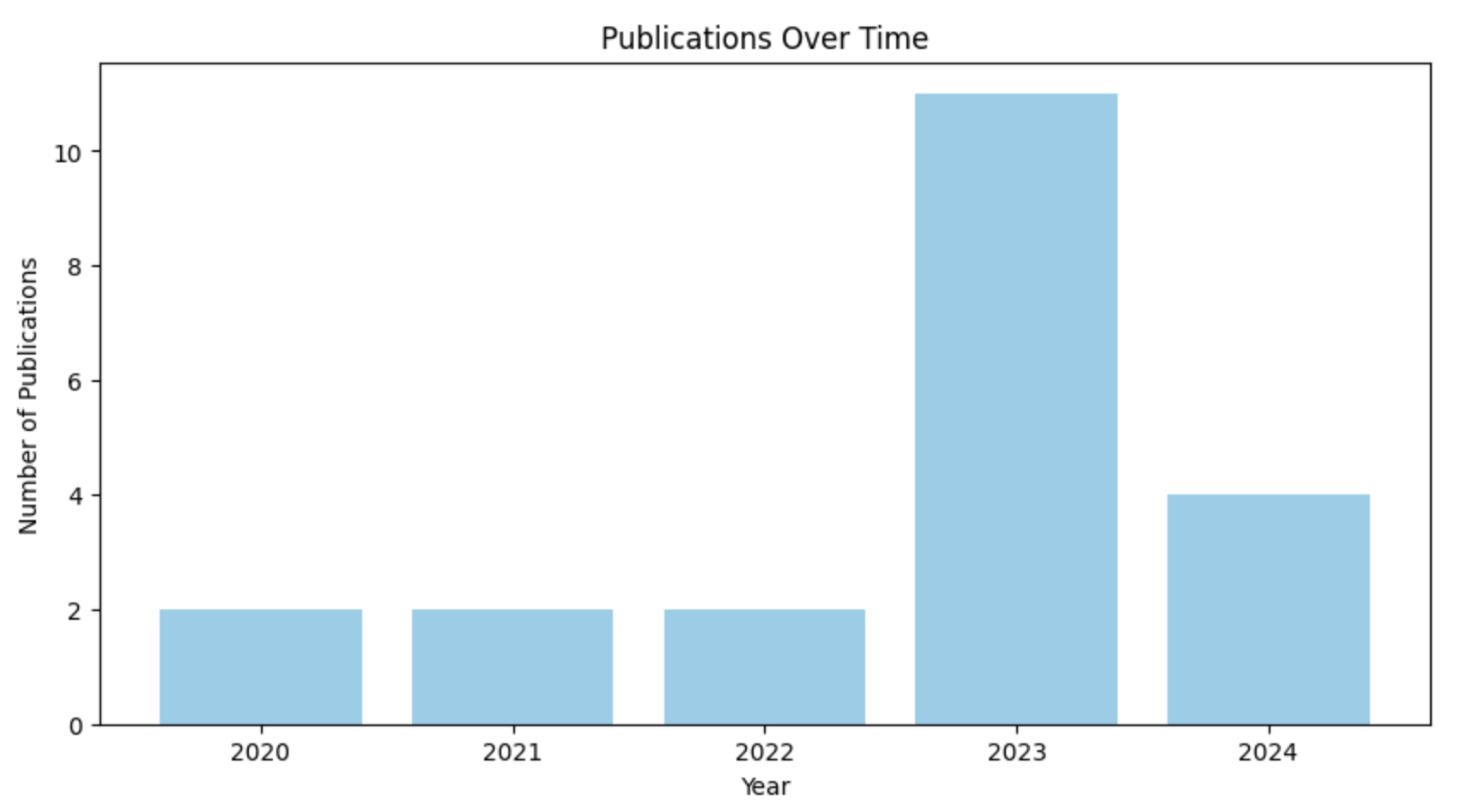}
  \caption{Publications of primary AR/VR papers over time.}
  \label{fig:publications}
\end{figure}

Although AR and VR concepts have existed for over three decades, significant security concerns only began to emerge about 15 years ago due to technological advancements. The first papers addressing security challenges in AR and VR systems were published in 2007 \cite{VRfirst,8}\cite{ARfirst,7}. As the use of these technologies increased, so did the importance of addressing their security concerns. Figure \ref{fig:publications} shows the number of papers selected after applying inclusion and exclusion criteria, highlighting a significant increase in publications in 2023. Since 2024 is not yet complete, the number of papers related to the vulnerabilities and weaknesses of these technologies for this year has not been fully recorded. Notably, all 5 papers that passed the quality assessment threshold were written in 2023.

\subsubsection{Significant Keyword Counts}
To identify common themes among the selected primary studies, we conducted a keyword analysis across all 5 studies. Table \ref{tab:keywords} shows the frequency of specific words appearing in these studies. Excluding keywords directly related to the topic, such as "VR" and "security," the analysis highlights key areas of focus in AR and VR security. For instance, the frequent mention of 'users' and 'user' underscores the importance of protecting user interactions and data. Keywords like 'https' and 'accessed' emphasize the critical nature of secure access and data transmission. The term 'information' reflects the crucial role of information handling in identifying and mitigating vulnerabilities in AR and VR systems.

\begin{table}[ht]
  \centering
  \caption{Frequency of specific keywords in the selected studies.}
  \label{tab:keywords}
  \begin{tabular}{|>{\centering\arraybackslash}m{3cm}|>{\centering\arraybackslash}m{2cm}|}
    \hline
    \textbf{Keyword} & \textbf{Frequency} \\
    \hline
    users & 216 \\
    \hline
    https & 215 \\
    \hline
    accessed & 163 \\
    \hline
    user & 136 \\
    \hline
    information & 127 \\
    \hline
    applications & 123 \\
    \hline
  \end{tabular}
\end{table}

\section{Findings}
Each primary study was examined to extract data about the initial steps of penetration testing, specifically reconnaissance and vulnerability assessment. The main findings are summarized in Table \ref{tab:findings}. Most studies cover both AR and VR technologies and provide detailed information about specific vulnerabilities and attacks, except for one paper that focuses solely on attacks. The subsequent sections categorize the identified vulnerabilities and attacks based on the focus areas of the papers: physical devices, development platforms, or applications.

\begin{table}[ht]
  \centering
  \caption{Main findings of the primary studies.}
  \label{tab:findings}
  \begin{tabular}{|>{\centering\arraybackslash}m{1cm}|>{\centering\arraybackslash}m{1cm}|>{\centering\arraybackslash}m{2cm}|>{\centering\arraybackslash}m{3cm}|}
    \hline
    \textbf{Primary Study} & \textbf{AR/VR} & \textbf{Attacks/ Vulnerabilities} & \textbf{Focus Area} \\
    \hline
    \cite{silva2023survey} & Both & Focus is on Attacks & Physical Devices \\
    \hline
    \cite{qamar2023systematic} & Both & Focus is on Attacks and Vulnerabilities & Physical Devices, Development Platforms and Applications/ Software/ Environments \\
    \hline
    \cite{2023metaverse} & Both & Focus is on Attacks and Vulnerabilities & Physical Devices, Development Platforms and Applications/ Software/ Environments \\
    \hline
    \cite{vondravcek2023rise} & VR & Focus is on Attacks and Vulnerabilities & Bigscreen Application \\
    \hline
    \cite{chandrashekar2023design} & VR & Focus is on Attacks and Vulnerabilities & Oculus Quest Device and MQTT Protocol \\
    \hline
    \cite{AR2023IOV} & AR & Focus is on Attacks and Vulnerabilities & Physical Devices, Development Platforms and Applications/ Software/ Environments \\
    \hline
  \end{tabular}
\end{table}

\section{Discussion}
\subsection{RQ1: What devices are used to implement AR and VR technology?}
In the paper "A Survey and Risk Assessment on Virtual and Augmented Reality Cyberattacks," several devices are highlighted as integral to AR and VR technology implementations. These devices include the Google Cardboard, Oculus Quest series (including the Oculus Quest 2), HTC Vive (including Vive Pro), and Microsoft's HoloLens. These devices are utilized across various sectors: gaming, healthcare, automotive, and education. In gaming, devices like the HTC Vive and Oculus Quest provide immersive experiences within secure, private networks at home. In healthcare, VR systems simulate environments for therapy under monitored conditions, ensuring privacy and security in networked institutional settings. In the automotive industry, VR facilitates efficient training without physical equipment, and in education, it enriches interactive learning experiences in more public network settings \cite{silva2023survey}.

The paper "A systematic threat analysis and defense strategies for the metaverse and extended reality systems" expands on the devices by discussing physical devices, development platforms, and applications/software/environments. Physical devices mentioned include the HTC Vive and Oculus Rift, which are fundamental in creating immersive experiences.\\
Regarding development platforms, Google's Cardboard SDK is an open-source programming platform for developing immersive experiences for Android and iOS. Node.js is widely used for developing 3D libraries and applications, making creating AR and VR content easier.
 ROBLOX is transitioning towards the Metaverse, allowing the creation of virtual games and experiences and even virtual shopping and business activities through its virtual Robux currency. SNAP.N (Snapchat) provides a platform for creating virtual avatars and AR filters, enabling the overlay of digital content onto the real world. The Unity Framework is extensively utilized for AR and VR software development, thanks to its comprehensive learning resources and compatibility with multiple operating systems. This makes it a key player in sectors such as television and film, virtual architecture, automotive design, and robotic simulations. Unreal Engine, another major platform, is renowned for its high-quality graphics and offers a marketplace filled with resources and educational materials.\\
Applications and environments assessed for vulnerabilities include popular Social Virtual Reality Learning Environment (VRLE) systems, the Bigscreen app, and Facebook Spaces, all of which play significant roles in providing immersive experiences \cite{9,qamar2023systematic}.

In the paper "Metaverse Cybersecurity Threats and Risks Analysis: The case of Virtual Reality Towards Security Testing and Guidance Framework," several physical devices integral to AR and VR technology are highlighted. These include the Oculus Quest, HTC Vive Pro, and Google Cardboard (utilizing smartphones such as the Nexus 6 and Galaxy S6). These devices are essential for delivering immersive virtual experiences by capturing and processing user interactions within virtual environments.\\
The OpenVR SDK is a significant development platform mentioned in the study. It is intended to allow cross-compatibility between various VR headset platforms, particularly trendy gadgets like the HTC Vive and Oculus Rift. This SDK is crucial for supporting various hardware devices and ensuring a seamless user experience across different VR systems.\\
The study also discusses several applications and environments used in AR and VR technology. These include BigScreen, VRChat, SteamVR, and High Fidelity. These applications provide users with interactive and immersive experiences, allowing them to engage in virtual chatrooms and environments. BigScreen, in particular, is highlighted for its immersive VR capabilities \cite{2023metaverse,8,2h}.

In the other paper "Rise of the Metaverse’s Immersive Virtual Reality Malware and the Man-in-the-Room Attack \& Defenses", researchers conducted an in-depth analysis of the Bigscreen VR application to understand its structure and functionalities better. The research began with the reconnaissance phase, where the authors gathered publicly available information about the Bigscreen system. They analyzed employment offers made by Bigscreen company to gain insights into the application's codebase and reviewed Bigscreen’s blog for updates and development posts. This analysis revealed various artifacts that remained on the disk drive by the Bigscreen application during earlier forensic investigations. It was found that the application’s UI elements were managed by JavaScript (with jQuery) in a limited integrated web browser, with bindings to the application’s core layer. This foundational knowledge provided the researchers with a comprehensive understanding of the application's inner workings. \\
Bigscreen is a VR platform designed for both entertainment and professional productivity. It supports activities such as playing computer games, watching movies, and remote collaboration. Users are represented by avatars that mimic their head and hand movements, creating an immersive experience. The application is available for Windows via Oculus Home, Steam, and Microsoft Store. Each virtual room in Bigscreen has a unique Room ID consisting of 8 alphanumeric characters. Rooms can be public or private (invite-only). Public rooms are listed on the application’s main screen, while private rooms can be joined using a confidential Room ID without further authorization. Bigscreen provides messaging capabilities for participants, with communication being peer-to-peer (P2P) and encrypted for security \cite{3h}.\\
To perform a detailed security analysis, the study set up a controlled laboratory environment. They used various tools, including mitmproxy, wifimitm, Scapy, and Wireshark, to analyze the application. The application’s encrypted traffic was decrypted using a temporary Certificate Authority (CA) configured on the VR workstations. Additionally, the researchers reverse-engineered and decompiled the application and its Dynamic Link Libraries (DLLs) into corresponding logic in C\#, revealing the inner structure of Bigscreen. The UI, implemented as a JavaScript environment communicating with the C\# core layer via JS-C\# function bindings, was thoroughly examined to identify potential vulnerabilities. This comprehensive approach allowed the researchers to gain a deep understanding of the application, setting the stage for identifying and exploiting its vulnerabilities \cite{vondravcek2023rise}.

The paper "Design \& Development of Virtual Reality Empowered Cyber-Security Training Testbed for IoT Systems" outlines a detailed simulation of a wastewater treatment facility using Virtual Reality (VR) and Internet of Things (IoT) technologies. The VR component, implemented using the Oculus Quest, provides an immersive environment where users can interact with a digital twin of the wastewater treatment plant. The Oculus Quest's wireless capabilities and robust processing power allow for a seamless, unrestricted experience. The VR interface simulates the entire water treatment process, from primary cleaning to final quality monitoring. Users can observe and adjust the virtual representation of the facility, providing a hands-on training environment that enhances understanding and operational efficiency.\\
The integration of IoT devices with the VR system is facilitated by the MQTT (Message Queuing Telemetry Transport) protocol, which enables real-time data exchange between physical hardware and the virtual environment. This setup includes various sensors, such as TDS and pH sensors, that monitor water quality parameters throughout the treatment process. The physical components, including pumps and mixers, are controlled via NodeMcu microcontrollers. By using MQTT for communication, the system ensures a dynamic and responsive simulation, allowing users to interact with the virtual plant in real-time \cite{chandrashekar2023design,10}.

In the paper "Secure Teleoperated Vehicles in Augmented Reality of Things: A Multichain and Digital Twin Approach," several AR technologies are discussed. The physical device introduced is the Augmented Reality-based Heads-Up Display (AR-HUD). AR-HUDs are integrated into vehicles and connected to various sensors to provide immersive overlays that enhance driver visibility and interaction. These overlays include AR navigation, lane violation alerts, and warnings for maintaining safe distances, among others. Originally derived from aviation technology, AR-HUDs project critical information directly onto the windshield, facilitating a safer and more interactive driving experience.\\
The development platform detailed in the paper is Multichain, a decentralized data security framework. Multichain addresses issues of latency and scalability by operating multiple blockchains in parallel. This platform secures volumetric video streams from AR-HUDs, ensuring data integrity and robustness against tampering during communication between teleoperated vehicles and their remote Digital Twins (DTs).\\
Regarding applications, environments, and software, the paper discusses Digital Twin (DT), which provides real-time monitoring and decision-making support for remote drivers, enhancing the effectiveness and safety of teleoperated driving. The Augmented Reality of Things (ARoT) integrates AR with the Internet of Things, enabling advanced functionalities like obstacle detection and augmented navigation through AR-HUDs. AR-HUDs are also part of Advanced Driver-Assistance Systems (ADAS), providing drivers with crucial information and alerts to prevent accidents and ensure safe driving. Furthermore, the AR-HUDs support volumetric video streaming with six Degrees of Freedom (DOF), which is essential for delivering high-quality, immersive experiences with minimal latency for remote teleportation \cite{AR2023IOV,11}.

\subsection{RQ2: What vulnerabilities are associated with these specific devices?}
According to the same paper, "A Survey and Risk Assessment on Virtual and Augmented Reality Cyberattacks," the vulnerabilities associated with these devices vary by the required connectivity and the nature of their use. Devices that require constant internet access, like those used in gaming and educational domains, are susceptible to a range of cyberattacks including Chaperone, Disorientation, Overlay, and Human Joystick attacks. This is due to the ease of convincing users to install malicious software. Conversely, in sectors like healthcare and automotive, where devices may not always be connected to the internet, the risk of cyberattacks such as Eavesdropping and Tampering still exists but is often due to physical access to the devices by unauthorized individuals. The automotive industry, with its restricted access environments, faces lower risks of network-based attacks but remains susceptible to internal threats like Impersonation and Jamming \cite{silva2023survey,h6}.

The paper "A systematic threat analysis and defense strategies for the metaverse and extended reality systems" identifies additional vulnerabilities. For example, research utilizing the Vivedump plugin on the HTC Vive headset demonstrated that the environments of users could be reconstructed from memory artifacts, presenting a notable privacy risk. Developers created an open-source plugin named Vivedump for performing memory forensics on metaverse systems, showcasing its application on the HTC Vive headset. The findings revealed that users’ immersive environments, including their locations, body postures, and room setups, could be reconstructed with ease from filtered memory artifacts using Vivedump’s features.\\
A vulnerability identified in the Google Cardboard SDK (CVE-2018-19111) allowed users’ personal and sensitive information to be exposed as unencrypted plaintext to a third-party website collecting Unity 3D statistics. Additionally, Node.js has encountered several exploits, including remote code execution, data manipulation, security bypasses, cross-site scripting (XSS), and memory exploitation.\\
ROBLOX has numerous security issues, such as cross-site scripting, weak hashing techniques, hardcoded API keys, and data leaks, which could potentially expose the sensitive information of nearly 100 million users. Researchers argue that these combined vulnerabilities could reveal names and email addresses, failing to meet basic CVSS security standards. SNAP.N (Snapchat) has experienced significant security flaws, including server-side request forgery, remote code execution, substantial authentication bypass, unrestricted file system access, and cross-site scripting. The Unity Framework has reported issues such as cross-site scripting and zero-day vulnerabilities, which allow attackers to get to locally saved documents, services, and web pages with the victim's authorization. Additional weaknesses in Unity can enable eavesdropping and man-in-the-room attacks during virtual sessions \cite{12}.\\
Unreal Engine has faced several security problems, including denial of service, data overwriting via directory traversal, and buffer overflow. Traditional cyber network attacks also impact immersive environments. For instance, various attacks took place towards prominent VRLE systems, including Syn flood, SQL injection, Ping sweeping, Intrusion, Eavesdropping, Packet flooding, Packet tampering, Packet Sniffing, Password attacks, Unauthorized logins, Information disclosure, and data tampering. Named attacks have a deleterious influence on user engagement in immersive environments, leading to cybersickness or motion sickness.\\
Eavesdropping is another vulnerability affecting both Bigscreen and VRLE systems, leading to privacy violations. Researchers have successfully conducted eavesdropping attacks in both Bigscreen and SocialVRLE applications. Forensic examinations of the Bigscreen app and Facebook Spaces, along with devices like the HTC Vive and Oculus Rift, revealed that attackers could mimic and visualize victims' immersive environments, alter metaverse sessions, and initiate man-in-the-middle (MITM) attacks due to unencrypted communication. The forensic study of Unity API and social networking metaverse applications like Bigscreen uncovered security gaps such as weaknesses in application input sanitization, enabling attackers to insert harmful code into the system. These unprotected input areas allowed the insertion of malicious JavaScript scripts, which ran on the internet browser and retrieved malware from web servers, impacting all users in the immersive environment. Multiple attacks were launched due to JavaScript execution and other flaws in the authentication method, including phishing, eavesdropping, unauthorized desktop sharing, man-in-the-room assault (MITR), and denial of service. VR phishing attacks, particularly in the Bigscreen app, are used to gain users' private and important data, such as credit card numbers, account details, and login passwords. In a particular investigation, scientists injected malware into users' PCs for phishing virtual sessions and created a replicating worm that infected every person involved in the virtual meeting \cite{qamar2023systematic,12}.

The paper "Metaverse Cybersecurity Threats and Risks Analysis: The case of Virtual Reality Towards Security Testing and Guidance Framework" explains various attacks on physical devices. Eavesdropping attacks have been demonstrated on the Oculus Quest, HTC Vive Pro, and Google Cardboard (using Nexus 6 and Galaxy S6). These attacks exploit facial vibrations captured by sensors to derive speech and user information. Malicious applications running on these VR platforms can analyze these vibrations to extract sensitive information, including voice commands, usage patterns, and lifestyle choices. This creates substantial privacy and security problems because these functions can be conducted without the user's consent.\\
Furthermore, an initial study demonstrated that an unsuspecting VR user may be taken to a physical site specified by the attacker. This kind of attack works by progressively modifying the user's virtual surroundings, prompting them to migrate in the real world to a specific area. The attack is the result of weaknesses in the OpenVR SDK, demonstrating the huge impact such flaws can have on the VR technology industry.\\
Several vulnerabilities have been uncovered in VR applications. Malicious actors, for example, can use vulnerabilities in the BigScreen VR application to spy on the victim's PC and obtain remote control of it. This assault also highlights how rapidly a worm can spread, infecting each user who enters a compromised room. Similar flaws were proven against VRChat, SteamVR, and High Fidelity, allowing attackers to change the victim's environment and run arbitrary code on the underlying operating system. Users are enticed into infected chatrooms, which trigger the exploit the moment they join \cite{2023metaverse}.

In the other paper "Rise of the Metaverse’s Immersive Virtual Reality Malware and the Man-in-the-Room Attack \& Defenses", the detailed security analysis of the Bigscreen application revealed several critical vulnerabilities that could be exploited for malicious purposes. One significant vulnerability was Remote Code Execution (RCE) via API Call to Application.OpenURL. The application was found to be vulnerable to RCE through this API call, which could be exploited to run arbitrary code on the victim's system. Another major issue identified was Cross-Site Scripting (XSS) vulnerabilities in user names, room names, room descriptions, and room categories, making the application susceptible to XSS attacks. Additionally, the application allowed DLLs to be patched without performing integrity checks, opening up the possibility of unauthorized modifications \cite{12}.\\
The research demonstrated various attacks and exploits that could be carried out against the Bigscreen application. One notable attack was the Man-in-the-Room (MitR) attack, where the attacker gained unauthorized access to a private VR room, moved around invisibly, and observed all activities without the victim’s knowledge or authorization. Another critical exploit was the VR worm, which illustrated how malware could spread among users sharing the same virtual room, leading to a widespread compromise of user privacy and security. The worm could modify the victim’s name to include an XSS payload, propagating the attack to other users and creating a botnet controlled by the attacker’s Command and Control (C\&C) server.\\
To carry out these attacks, the study followed a structured methodology. In Phase I, the reconnaissance phase, the researchers examined Bigscreen and gathered publicly available information, including job offers and blog posts. In Phase II, the laboratory setup and tool sets phase, they prepared laboratory equipment and software tools based on identified areas of interest. During Phase III, the security analysis phase, they evaluated the general security by analyzing network traffic, conducting penetration testing, and reverse engineering the protocols and Bigscreen desktop program. Phase IV focused on exploit development, where they crafted exploits to evaluate the effects of identified vulnerabilities. In Phase V, tool creation, they combined known attacks and exploits to create an extensive attack tool. Finally, Phase VI involved testing, where they evaluated the success rate according to defined scenarios, summarized results, and responsibly disclosed findings to vendors.\\
The testing phase yielded significant results, demonstrating the crucial nature of the found vulnerabilities in real-world circumstances. All tested attacks needed no input from the victim, resulting in zero-click attacks. The researchers successfully executed attacks in all scenarios, including passive stay in the lobby, creation of public and private rooms, conducting private meetings, and transitioning between rooms. The Man-in-the-Room attack allowed the attacker to gain unauthorized access to private VR rooms and remain invisible while eavesdropping on the victims. The VR worm demonstrated how malware could spread across the entire Bigscreen community, leading to a widespread compromise of user privacy and security.\\
The study's findings underscored the importance of robust security measures in VR applications to protect users from potential threats. The detailed examination of Bigscreen’s vulnerabilities and the demonstration of various attacks provided valuable insights for improving the security posture of VR platforms.

The paper "Design \& Development of Virtual Reality Empowered Cyber-Security Training Testbed for IoT Systems" identifies several critical vulnerabilities associated with the VR system used in the context of an IoT-based wastewater treatment facility. One significant vulnerability is the use of outdated software and firmware in the VR system, including the Oculus Quest. These outdated components are no longer supported by vendors, making them susceptible to known vulnerabilities that attackers can exploit to gain unauthorized access or disrupt operations. This issue is particularly concerning given the critical nature of the wastewater treatment process, where any disruption can have severe consequences.\\
Another major vulnerability is the lack of proper authentication and authorization mechanisms in the VR system's integration with IoT devices. The communication between the VR system and IoT devices relies on protocols such as MQTT. Without robust authentication and authorization measures, attackers could potentially intercept or manipulate the data being transmitted, leading to unauthorized control of the system. This could allow an attacker to alter the functioning of the wastewater treatment plant, resulting in significant operational and safety issues \cite{h5}.\\
Additionally, the paper highlights the potential weakness in encryption practices. If the communication between the VR system and IoT devices is not adequately encrypted, it opens the door for attackers to eavesdrop on the data exchange. Weak or improperly implemented encryption can allow attackers to capture sensitive information and use it to exploit the system further. Ensuring strong encryption practices are in place is crucial to protect the integrity and confidentiality of the data being transmitted between the VR system and IoT devices. These identified vulnerabilities underscore the importance of implementing comprehensive security measures to safeguard VR systems used in critical infrastructure settings like wastewater treatment facilities \cite{chandrashekar2023design}.

In the paper "Secure Teleoperated Vehicles in Augmented Reality of Things: A Multichain and Digital Twin Approach," several vulnerabilities and attack vectors related to AR technologies are discussed in detail. AR-HUDs, being a central component of the teleoperated vehicle's AR system, are susceptible to security vulnerabilities, particularly concerning the integrity of the data displayed. One significant vulnerability is the risk of false data injection attacks, where attackers can manipulate the data being projected onto the AR-HUD. The paper highlights the potential for forged frames in the AR-HUD video streams shared between teleoperated vehicles and their Digital Twins (DTs). This forged data can mislead drivers and cause dangerous driving situations. At the network layer, adversaries can falsify communicated data using various strategies, including man-in-the-middle attacks, Denial of Service (DoS) attacks, and cloning of data. The paper identifies the man-in-the-middle attack as the most common and stealthy method. This attack involves tampering with frames of the AR-HUD video during transmission, resulting in altered video displays on the DT and falsified AR embeddings. The impact of such attacks is illustrated through a comparison of the original AR-HUD video frame with a forged frame, showing how the video and AR annotations can be manipulated to deceive the system and the driver.
To address these vulnerabilities, the paper proposes a Multichain framework for securing the data. Multichain is designed to handle issues of latency and scalability by operating multiple blockchains in parallel, ensuring that the volumetric video streams from AR-HUDs are secure. This framework prevents tampering and falsification of data during transmission, ensuring the integrity and authenticity of the AR-HUD frames.\\
The DTs used in this system for real-time monitoring and decision-making are also targets for potential attacks. If the AR-HUD video data transmitted to the DTs is compromised, it can lead to incorrect decisions by the remote driver, potentially causing accidents.
The ARoT framework integrates AR with IoT, introducing additional vulnerabilities through interconnected devices. These devices must be secure to prevent unauthorized access and data manipulation, which could compromise the entire ARoT system.
ADAS relies heavily on accurate data provided by AR-HUDs. Any attack that injects false data into these systems can result in failures of critical safety features such as collision warnings and lane departure alerts.
Volumetric video streaming used by AR-HUDs introduces challenges in maintaining data integrity and low latency. The paper addresses this by using Multichain to ensure the secure and timely transmission of video data. However, vulnerabilities such as man-in-the-middle attacks during transmission could still pose significant risks if not adequately mitigated.
The proposed Multichain framework aims to secure AR-HUDs and their associated data streams against these vulnerabilities, providing a robust defense against tampering and ensuring the safety and effectiveness of teleoperated driving systems \cite{AR2023IOV}.

These findings underscore the extensive attack surface and the critical need for comprehensive security measures in AR and VR systems. For a summarized breakdown of these vulnerabilities, please refer to Tables \ref{tab:physical_device}, \ref{tab:development_platforms}, and \ref{tab:applications_software}.

\begin{table}[ht]
\centering
\caption{Attacks/vulnerabilities based on physical device}
\label{tab:physical_device}
\begin{tabular}{|>{\centering\arraybackslash}m{2cm}|>{\centering\arraybackslash}m{1cm}|>{\centering\arraybackslash}m{4.5cm}|}
\hline
\textbf{Headset} & \textbf{AR/VR} & \textbf{Attacks} \\ \hline
All headsets & VR/AR & Observation \cite{silva2023survey} \\ \hline
Google Cardboard & AR/VR & Eavesdropping, Side-channel \cite{silva2023survey}, \cite{2023metaverse} \\ \hline
HTC Vive & VR & Chaperone, Disorientation, Overlay, Human Joystick, Unauthorized access, Tampering, Impersonation, Hijacking, Jamming, MITM, Memory dump analysis leading to mimicry and visualization of environments, Integrity violations, Content tampering, Obstruction of users' views \cite{silva2023survey} \cite{qamar2023systematic}\\ \hline
HTC Vive & AR/VR & DOS, Unauthorized access, Observation \cite{silva2023survey} \\ \hline
HTC Vive Pro & AR/VR & Eavesdropping, Side-channel \cite{silva2023survey}, \cite{2023metaverse} \\ \hline
HTC Vive VR & VR/AR & Hijacking \cite{silva2023survey} \\ \hline
HTC Vive VR & VR & Jamming \cite{silva2023survey} \\ \hline
Oculus Quest & AR/VR & Eavesdropping, Side-channel \cite{silva2023survey}, \cite{2023metaverse} \\ \hline
Oculus Quest & VR & Run malicious code \cite{silva2023survey} \\ \hline
Oculus Quest 2 & VR & Ransomware \cite{silva2023survey} \\ \hline
Oculus Rift & VR & MITM, Memory dump analysis leading to mimicry and visualization of environments, Integrity violations, Content tampering, Obstruction of users' views \cite{qamar2023systematic} \\ \hline
AR-HUDs & AR & MITM, False data injection, Tampering with AR-HUD video frames during transmission \cite{AR2023IOV} \\ \hline
\end{tabular}
\end{table}

\begin{table}[ht]
\centering
\caption{Attacks/vulnerabilities based on development platforms}
\label{tab:development_platforms}
\begin{tabular}{|>{\centering\arraybackslash}m{2cm}|>{\centering\arraybackslash}m{1cm}|>{\centering\arraybackslash}m{4.5cm}|}
\hline
\textbf{Name} & \textbf{AR/VR} & \textbf{Attacks/Vulnerabilities} \\ \hline
Google’s Cardboard SDK & VR & CVE-2018-19111 \cite{qamar2023systematic} \\ \hline
Node.js & AR/VR & Remote code execution, manipulation of data, bypassing of security checks, XSS, memory exploitation \cite{qamar2023systematic} \\ \hline
MQTT Protocol & AR/VR & Vulnerabilities in MQTT protocol without proper encryption, Data manipulation, Network spying, Authentication flaws, Authorization issues \cite{chandrashekar2023design} \\ \hline
Roblox & VR & Cross-site scripting, inappropriate hashing algorithms, hardcoded API keys, Janus vulnerability, data exposure \cite{qamar2023systematic} \\ \hline
SNAP.N & AR & Server-side request forgery, server-side remote code execution, authentication bypass, unrestricted file system access, XSS, remote freezing of mobile phones \cite{qamar2023systematic} \\ \hline
Unity Framework & AR/VR & Cross-site scripting, zero-day vulnerabilities, out-of-bounds memory leaks, input string validation exploits, eavesdropping, man-in-the-room \cite{qamar2023systematic} \\ \hline
Unreal Engine & AR/VR & Denial of service, overwriting of data by directory traversal, buffer overflow \cite{qamar2023systematic} \\ \hline
OpenVR SDK & VR & Virtual Environment Manipulation \cite{2023metaverse} \\ \hline
\end{tabular}
\end{table}

\begin{table}[ht]
\centering
\caption{Attacks/vulnerabilities based on applications/software/environments}
\label{tab:applications_software}
\begin{tabular}{|>{\centering\arraybackslash}m{2cm}|>{\centering\arraybackslash}m{1cm}|>{\centering\arraybackslash}m{4.5cm}|}
\hline
\textbf{Name} & \textbf{AR/VR} & \textbf{Attacks/Vulnerabilities} \\ \hline
Bigscreen VR Application & VR & Application input sanitization flaws, Injection of malicious code, JavaScript execution, Malware download, Phishing, Eavesdropping, Illegitimate desktop sharing, Man-in-the-room attack, Denial of services, Remote control, Arbitrary code execution, Environment alteration, Remote Code Execution (RCE) via `Application.OpenURL`, XSS vulnerabilities in user inputs, Patching DLL without integrity check, Lack of integrity in WebRTC, Lack of authentication in signaling channel, Botnet and VR worm spreading, MITM, Memory dump analysis leading to mimicry and visualization of environments, Integrity violations, Content tampering, Obstruction of users' views \cite{qamar2023systematic}, \cite{2023metaverse}, \cite{vondravcek2023rise} \\ \hline
Wastewater Treatment Plant Simulator & VR & Unauthorized access to IoT devices, Cyber intrusions (DoS, phishing, unpatched systems), Vulnerabilities in MQTT protocol \cite{chandrashekar2023design} \\ \hline
VRChat & VR & Arbitrary code execution, Environment alteration \cite{2023metaverse} \\ \hline
SteamVR & VR & Arbitrary code execution, Environment alteration \cite{2023metaverse} \\ \hline
High Fidelity & VR & Arbitrary code execution, Environment alteration \cite{2023metaverse} \\ \hline
VRLE (Social Virtual Reality Learning Environment) systems & VR & Syn flood, SQL injection, Ping sweeping, Intrusion, Eavesdropping, Packet flooding, Packet tampering, Packet Sniffing, Password attacks, Unauthorized logins, Information disclosure, DOS, Data leakage \cite{qamar2023systematic} \\ \hline
Facebook Spaces & VR & MITM, Memory dump analysis leading to mimicry and visualization of environments, Integrity violations, Content tampering, Obstruction of users' views \cite{qamar2023systematic} \\ \hline
Digital Twin & AR & Potential attacks on transmitted AR-HUD video data, Incorrect decision-making due to compromised data \cite{AR2023IOV} \\ \hline
Augmented Reality of Things (ARoT) & AR & Unauthorized access and data manipulation \cite{AR2023IOV} \\ \hline
Advanced Driver-Assistance Systems (ADAS) & AR & Vulnerabilities due to false data injection and Failure of critical safety features (e.g., collision warnings, lane departure alerts) \cite{AR2023IOV} \\ \hline
Volumetric Video Streaming & AR & Integrity and latency issues, Susceptibility to MITM attacks during transmission \cite{AR2023IOV} \\ \hline
\end{tabular}
\end{table}

\section{Conclusion}
This paper presents a systematic review, emphasizing security challenges with VR and AR technologies. Specifically, the paper focuses on the reconnaissance phase and the assessment of vulnerabilities as an initial step in penetration testing. Our findings point out the severe security threats that these technologies pose. The review revealed device and platform vulnerabilities, such as overlay attacks, human joystick attacks, and tampering, which appear highly severe to end users.\\
The reconnaissance phase in this study considered and categorized VR and AR technologies into three sub-categories: physical devices, development platforms, and applications/environments. Physical devices such as the HTC Vive, Oculus Quest, and Google Cardboard were found to be vulnerable to attacks like eavesdropping, side-channel attacks, and unauthorized access. Development platforms, including Unity and Unreal Engine, were identified with vulnerabilities like cross-site scripting and remote code execution. Applications such as Bigscreen VR and VRChat exhibited severe security issues, including remote code execution, man-in-the-room attacks, and malware propagation.\\
Among the most repeated technologies, the HTC Vive and Oculus Quest stood out due to their wide usage and the variety of attacks they faced, such as overlay attacks and human joystick attacks, which exploit the devices’ functionalities to manipulate user environments. Repeated vulnerabilities across these devices and platforms included cross-site scripting, remote code execution, and eavesdropping, indicating common security weaknesses that need to be addressed.\\
In summary, while AR and VR technologies offer significant potential across various industries, their integration introduces complex security challenges that require robust measures. This paper's identification and categorization of vulnerabilities provide a critical foundation for future studies aimed at enhancing the security of AR and VR systems. The findings emphasize the need for continuous monitoring and updating of security protocols to mitigate emerging threats, ensuring the safe deployment and use of these transformative technologies.

\section{Future Work}
This study provides a detailed overview of AR and VR security, yet further investigation is essential to address the evolving landscape of these technologies. Future studies should consider a more extended timeframe beyond the 2020-2024 period. Analyzing older publications can provide historical context and highlight how vulnerabilities and attack vectors have evolved. Additionally, including recent advancements post-2024 will ensure the study remains current and relevant.\\
Expanding the range of databases and sources can provide a more comprehensive view of the security landscape. Including sources such as ACM Digital Library, SpringerLink, and Google Scholar can uncover additional insights and research that may have been missed.\\
While the focus on reconnaissance and vulnerability assessment phases is crucial, future research should also explore the subsequent steps in penetration testing, including exploitation, post-exploitation, and reporting. This comprehensive approach will offer a complete understanding of how vulnerabilities are identified, exploited, and addressed. Identifying and testing current exploitation techniques tailored to AR and VR devices will validate vulnerabilities and assess their real-world impact. Additionally, developing effective remediation strategies, including targeted patches and security frameworks, is crucial for enhancing AR and VR security. Ensuring robust and adaptive security measures will be vital as AR and VR technologies continue to advance and integrate into various sectors.


\bibliographystyle{IEEEtran}
\clearpage
\bibliography{references}

\end{document}